\documentstyle{mn}
\newif\ifAMStwofonts
\ifoldfss
  \ifCUPmtlplainloaded \else
    \NewTextAlphabet{textbfit} {cmbxti10} {}
    \NewTextAlphabet{textbfss} {cmssbx10} {}
    \NewMathAlphabet{mathbfit} {cmbxti10} {} 
    \NewMathAlphabet{mathbfss} {cmssbx10} {} 
  \fi
  \ifAMStwofonts
    \ifCUPmtlplainloaded \else
      \NewSymbolFont{upmath} {eurm10}
      \NewSymbolFont{AMSa} {msam10}
      \NewMathSymbol{\upi}     {0}{upmath}{19}
      \NewMathSymbol{\umu}     {0}{upmath}{16}
      \NewMathSymbol{\upartial}{0}{upmath}{40}
      \NewMathSymbol{\leqslant}{3}{AMSa}{36}
      \NewMathSymbol{\geqslant}{3}{AMSa}{3E}

    \fi
  \fi
\fi 

\ifnfssone
  \newmathalphabet{\mathit}
  \addtoversion{normal}{\mathit}{cmr}{m}{it}
  \addtoversion{bold}{\mathit}{cmr}{bx}{it}
  \newmathalphabet{\mathbfit} 
  \addtoversion{normal}{\mathbfit}{cmr}{bx}{it}
  \addtoversion{bold}{\mathbfit}{cmr}{bx}{it}
  \newmathalphabet{\mathbfss} 
  \addtoversion{normal}{\mathbfss}{cmss}{bx}{n}
  \addtoversion{bold}{\mathbfss}{cmss}{bx}{n}
  \ifAMStwofonts
    \ifCUPmtlplainloaded \else
      \UseAMStwoboldmath
      \makeatletter
      \new@mathgroup\upmath@group
      \define@mathgroup\mv@normal\upmath@group{eur}{m}{n}
      \define@mathgroup\mv@bold\upmath@group{eur}{b}{n}
      \edef\UPM{\hexnumber\upmath@group}
      \new@mathgroup\amsa@group
      \define@mathgroup\mv@normal\amsa@group{msa}{m}{n}
      \define@mathgroup\mv@bold\amsa@group{msa}{m}{n}
      \edef\AMSa{\hexnumber\amsa@group}
      \makeatother
      \mathchardef\upi="0\UPM19
      \mathchardef\umu="0\UPM16
      \mathchardef\upartial="0\UPM40
      \mathchardef\leqslant="3\AMSa36
      \mathchardef\geqslant="3\AMSa3E
    \fi
  \fi
\fi 

\ifnfsstwo
  \DeclareMathAlphabet{\mathbfit}{OT1}{cmr}{bx}{it}
  \SetMathAlphabet\mathbfit{bold}{OT1}{cmr}{bx}{it}
  \DeclareMathAlphabet{\mathbfss}{OT1}{cmss}{bx}{n}
  \SetMathAlphabet\mathbfss{bold}{OT1}{cmss}{bx}{n}
  \ifAMStwofonts
    \ifCUPmtlplainloaded \else
      \DeclareSymbolFont{UPM}{U}{eur}{m}{n}
      \SetSymbolFont{UPM}{bold}{U}{eur}{b}{n}
      \DeclareSymbolFont{AMSa}{U}{msa}{m}{n}
      \DeclareMathSymbol{\upi}{0}{UPM}{"19}
      \DeclareMathSymbol{\umu}{0}{UPM}{"16}
      \DeclareMathSymbol{\upartial}{0}{UPM}{"40}
      \DeclareMathSymbol{\leqslant}{3}{AMSa}{"36}
      \DeclareMathSymbol{\geqslant}{3}{AMSa}{"3E}
    \fi
  \fi
\fi 

\ifCUPmtlplainloaded \else
  \ifAMStwofonts \else 
    \def\upi{\pi}
    \def\umu{\mu}
    \def\upartial{\partial}
  \fi
\fi

\title[Upper limits on the central black hole masses of 47Tuc and
NGC6397]{Upper limits on the central black hole
masses of 47Tuc and NGC6397 from radio continuum emission} \author[S. De
Rijcke, P. Buyle, H. Dejonghe] {S. De Rijcke$^1$ \thanks{Postdoctoral Fellow of the
Fund for Scientific Research - Flanders
(Belgium)(F.W.O)}\thanks{corresponding author: sven.derijcke@UGent.be}
\& P. Buyle$^1$, H. Dejonghe$^1$ \\ $^1$ Sterrenkundig Observatorium, Universiteit
Gent, Krijgslaan 281, S9, B-9000, Gent, Belgium} \date{Accepted 1988
December 15.  Received 1988 December 14; in original form 1988 October
11}

\pagerange{\pageref{firstpage}--\pageref{lastpage}}
\pubyear{1994}

\begin{document}

\maketitle

\label{firstpage}

\begin{abstract}
We present upper-limits on the masses of the putative central
intermediate-mass black holes in two nearby Galactic globular
clusters:~47Tuc (NGC104), the second brightest Galactic globular
cluster, and NGC6397, a core-collapse globular cluster and, with a
distance of 2.7~kpc, quite possibly the nearest globular cluster,
using a technique suggested by T. Maccarone. These mass estimates have
been derived from 3$\sigma$ upper limits on the radio continuum flux
at 1.4~GHz, assuming that the putative central black hole accretes the
surrounding matter at a rate between 0.1~\% and 1~\% of the Bondi
accretion rate. For 47Tuc, we find a 3$\sigma$ upper limit of $2060 -
670 M_\odot$, depending on the actual accretion rate of the black hole
and the distance to 47Tuc. For NGC6397, which is closer to us, we
derive a 3$\sigma$ upper limit of $1290 - 390 M_\odot$. While
estimating mass upper-limits based on radio continuum observations
requires making assumptions about the gas density and the accretion
rate of the black hole, their derivation does not require complex and
time consuming dynamical modeling. Thus, this method offers an
independent way of estimating black hole masses in nearby globular
clusters. If, generally, central black holes in stellar systems
accrete matter faster than 0.1~\% of the Bondi accretion rate, then
these results indicate the absence of black holes in these globular
clusters with masses as predicted by the extrapolated $M_\bullet -
\sigma_{\rm c}$ relation.
\end{abstract}


\section{Introduction}

Theory provides us with different ways of producing intermediate-mass
black holes (IMBHs) in dense stellar clusters, such as globular
clusters. IMBHs are defined here as having a mass larger than that of
stellar black holes but smaller than that of the super-massive black
holes (SMBH) found in galaxies, so $10^2\, M_\odot < M_\bullet < 10^6
\, M_\odot$. One scenario envisages an IMBH to grow from a
stellar-mass seed black hole, the remnant of a massive star that
exploded as a supernova \cite{mi02}. This seed black hole rapidly
sinks to the cluster center due to dynamical friction, or, in the case
of a post-core-collapse cluster, was formed near the cluster center
since that is where the most massive stars can be found, and grows by
accreting stars, less massive stellar black holes or interstellar
gas. In a core-collapse cluster, stars near the core are packed so
densely that some may collide, causing a runaway growth of the
remnant's mass \cite{pz02}. Both mechanisms seem to lead naturally to
IMBHs with a mass that is about 0.1\% of the cluster mass. The IMBH
can be even more massive (up to $10^4 M_\odot$) if it originated from
multiple seeds. Another suggested way of forming IMBHs considers the
radiation drag that is exerted by stars on the interstellar medium
\cite{ka03,no05}. Radiation drag is expected to funnel gas, expelled
by SN{\sc ii} and SN{\sc i}a explosions, towards the cluster center
where it is likely to form a massive central object surrounded by an
accretion disk. It is expected that this central object will collapse
into an IMBH if its mass exceeds a few 100 $M_\odot$. In massive
galaxies, this process may lead to central black holes that have a
mass of about 0.1\% of the bulge mass \cite{ka03}. However, in the
case of globular clusters, IMBHs formed this way are expected to be
much less massive than 0.1\% of the cluster mass (the ratio of the
IMBH mass to the cluster mass is rather of the order of $10^{-5}$). In
fact, clusters less massive than $\sim 5 \times 10^6 M_\odot$ are not
expected to contain a central black hole, although they can harbor a
central object not massive enough to collapse into a black hole
\cite{no05}. Yet another, somewhat more speculative, possibility is
that Population {\sc iii} stars, with masses of a few hundred solar
masses, were the seeds of IMBHs \cite{ma01}. Nuclear energy production
in Population {\sc iii} stars more massive than $\sim 260 M_\odot$
cannot counteract gravity and such objects immediately collapse into
black holes. Such very massive seed black holes, which could be called
IMBHs in their own right, then grow further by accretion. This way of
producing IMBHs, unlike the previous ones, does not require the host
stellar system to be dense.

Therefore, even upper limits on the masses of central black holes in
globular clusters are of great interest, since they constrain the
possible formation scenarios for IMBHs and the nature of IMBH
progenitors. In this paper, we present upper limits on the IMBH masses
of 47Tuc and NGC6397, two of the nearest globular clusters, derived
from 20~cm continuum observations, obtained with the Australia
Telescope Compact Array (ATCA). In Section \ref{detect}, we briefly
discuss how IMBHs can be detected via their radio continuum
emission. In Sect. \ref{obs}, the observations and the data reduction
procedures are presented, followed by a discussion of the results in
Sect. \ref{disc}. We summarize our conclusions in section \ref{conc}.

\section{Detecting IMBHs using radio continuum emission} \label{detect}

While various approaches to the estimation of IMBH masses in globular
clusters have been suggested and/or applied (such as dynamical
modeling of stellar kinematics, see e.g. Gerssen et
al. \shortcite{ge02}, or the detection of high-proper-motion stars
near the globular cluster center \cite{dr03}) one of the
observationally least demanding methods is the possibility to detect
black holes through the radio emission they produce. Gas that is
accreted onto a black hole converts part of its rest mass into
radiation which, in the mass regime of IMBHs, is most easily recovered
at radio wavelengths. The equation that relates the mass of a black
hole, $M_\bullet$, in a stellar system at a distance $d$ to its X-ray
luminosity $L_X$ and radio continuum emission $F_{\rm cont}$ reads as
\begin{eqnarray}
F_{\rm cont} &=& 10 \left( \frac{L_X}{3 \times 10^{31} \,{\rm erg s}^{-1}}
\right)^{0.6} \left( \frac{M_\bullet}{100 \,M_\odot} \right)^{0.78} \nonumber \\
&& \times \left( \frac{d}{10 \,{\rm kpc}} \right)^{-2}\, \mu{\rm Jy},
\end{eqnarray}
assuming a flat radio spectrum \cite{ma04} and the Fundamental-Plane
relationship between radio luminosity, X-ray luminosity, and black
hole mass \cite{mhd03,fa04}. For a given accretion rate, which is
usually expressed as a fraction of the Bondi accretion rate
\begin{eqnarray}
\dot{M}_{\bullet,\rm Bondi} &=& 3.2 \times 10^{14} \left(
\frac{M_\bullet}{2000\,M_\odot} \right)^2 \left( \frac{n}{0.2 \, {\rm
cm}^{-3}} \right) \nonumber \\ &&  \times \left( \frac{T}{10^4\,{\rm K}}
\right)^{-1.5} \, {\rm kg \, s}^{-1},
\end{eqnarray}
(with $n$ and $T$ the gas density and temperature, respectively) the
X-ray luminosity can be calculated, assuming that 10~\% of the rest
mass energy of the infalling matter is converted into radiation, and
that the total luminosity $L_{\rm tot}$ is related to the X-ray
luminosity as given by $L_{\rm tot} = 6\times 10^{-3} L_X^{0.5} + L_X$
(using the Eddington luminosity as unit). This relation takes into
account the energy of a relativistic jet \cite{fe03}. This way, the
radio continuum flux $F_{\rm cont}$ can be expressed as a function of
the black hole mass $M_\bullet$, the distance $d$, and the gas density
$n$ and temperature $T$. In the following, we will assume $n =
0.067$~cm$^{-3}$ and $T = 10^4$~K (see Freire et al. \shortcite{fr01})
as typical values.

As shown by e.g. Di Matteo et al. \shortcite{di01} and Quataert \&
Gruzinov \shortcite{qg00} for nearby early-type galaxies and for the
Milky Way, respectively, massive central black holes seem to accrete
much slower than predicted by the Bondi estimate (with
$\dot{M}_\bullet < 10^{-2} \dot{M}_{\bullet,\rm Bondi}$). The low
number of Galactic neutron stars that were detected in the ROSAT
All-Sky Survey can be explained if they are accreting at a rate of
roughly 0.1~\% of the Bondi rate \cite{pe03}, although, of course, a
neutron star accreting from the interstellar medium is a far cry from
a central black hole. Theoretical models of the accretion of matter
that has non-zero angular momentum onto a black hole \cite{kr05,pb03}
show a reduction of the accretion rate down to the level of a few
percent of the Bondi rate. In the following, we will assume an
accretion rate in the range $ 10^{-3} \dot{M}_{\bullet,\rm Bondi} <
\dot{M}_\bullet < 10^{-2} \dot{M}_{\bullet,\rm Bondi}$, in agreement
with Maccarone et al. \shortcite{ma04} and Maccarone et
al. \shortcite{ma05}. Thus, it is possible to estimate the masses (or
place upper limits thereon) of central black holes in nearby globular
clusters using straightforward and relatively short radio
observations.

\section{Observations and data reduction} \label{obs}

We observed both NGC6397 and 47Tuc with the ATCA on respectively 2004
December 23 and 26. We used the 1.5D configuration, with baselines
ranging from 107 to 4439~m. The continuum observations were made at a
central frequency of 1.384~GHz in the FULL\_128\_1 correlator
configuration with a bandwidth of 128~MHz divided over 32 channels
into 4 products. At the start of each observation, we observed the
standard ATCA primary calibrator 1934-638 for 10~min. A secondary
calibrator was observed every 45~min. The total integration time
(including the calibration) for each globular cluster was 3.5~h. The
data reduction was performed with the MIRIAD package \cite{stw95}, the
standard ATCA data analysis program. The observed data were loaded
into MIRIAD with the birdie, xycorr, and reweight options, which
respectively flags out the channels that are affected by the ATCA
self-interference, corrects for the phase difference between the $X$
and $Y$ channels and re-weights the visibility spectrum in the lag
domain to eliminate the effects of the Gibbs phenomena. This results
in 13 channels, each 4~MHz wide. The usual data reduction steps,
including phase, amplitude and bandpass calibration, were afterwards
performed. Instead of applying the traditional CLEAN algorithm
\cite{ccw90} we used the Multi-Frequency Synthesis (MFS) method
\cite{sc99} with a natural weighting to surpress the noise. This
improved drastically the ($u,v$)-coverage by eliminating spectral
artifacts.

For NGC6397, after applying all above reduction steps, we constructed
a radio continuum map with a beam size of $36.08''\times 15.64''$ and
a 1$\sigma$ noise value of 72~$\mu$Jy. No radio continuum sources could be
detected in this object at this noise level. We therefore assume
216~$\mu$Jy as the 3$\sigma$ upper limit on the radio emission of the
black hole.  

Our 20~cm map of 47Tuc has a depth similar to that presented by
McConnell et al. \shortcite{mc04} although it has higher spatial
resolution : our final image has a beam size of $33.39''\times
12.55''$ and a 1$\sigma$ noise value of 75~$\mu$Jy. None of the radio
continuum sources is close enough to the cluster center to be
associated with a putative black hole. We therefore assume 225~$\mu$Jy
as the 3$\sigma$ upper limit on the radio emission of the black hole.

\section{Results} \label{disc}

47Tuc is a massive globular cluster, with a total stellar mass $M
\approx 1.5 \times 10^6 M_\odot$ \cite{ge95}. Its large core radius
suggests that it is a relaxed (and not a post-core-collapse) cluster
\cite{ho00}. We use the distance modulus $m-M = 13.50 \pm 0.08$~mag
derived by Gratton et al. \shortcite{gr03}. NGC6397 on the other hand
is a much less massive cluster, with a total stellar mass $M \approx
1.0 \times 10^5 M_\odot$ \cite{m91}. It is a very centrally
concentrated, collapsed cluster. Using accurate color-magnitude
diagrams and luminosity functions of two fields inside NGC6397,
Andreuzzi et al. \shortcite{a04} found strong evidence for different
mass distributions and hence for mass segregation to have occurred. We
adopt a distance modulus $m-M = 12.13 \pm 0.15$~mag \cite{rg98},
determined by main-sequence fitting using lower main sequence stars
with {\tt hipparcos} parallaxes measured to a precision of better than
10\%.

\begin{figure}
\vspace*{8.0cm}
\special{hscale=92 vscale=92 hsize=500 vsize=500
hoffset=-18 voffset=-330 angle=0 psfile="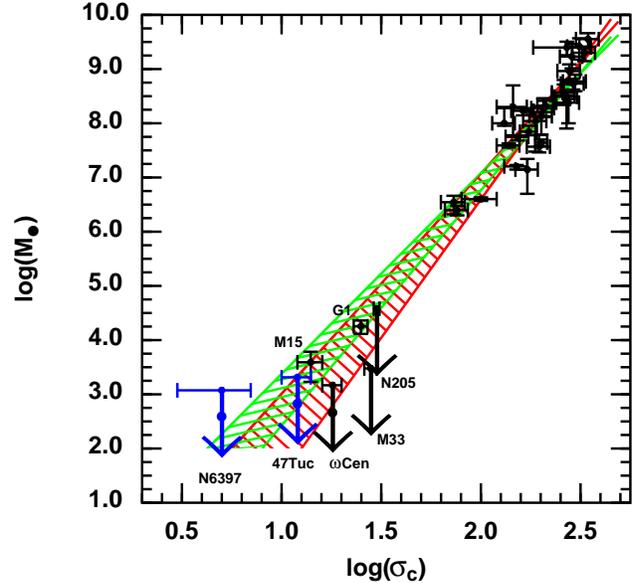"}
\caption{The $M_\bullet - \sigma_{\rm c}$ relation of massive
elliptical and spiral galaxies, dwarf galaxies, and globular
clusters. The $M_\bullet - \sigma_{\rm c}$ derived by Ferrarese \&
Ford (2005) lies within the red region, which quantifies the 1$\sigma$
uncertainty on that relation. The relation and its 1$\sigma$
uncertainty as derived by Tremaine et al. (2002) is indicated in
green. The blue datapoints correspond to the 3$\sigma$ upper limits on
the masses of the black holes in 47Tuc and NGC6397, the upper dot
corresponding to an accretion rate equal to $\dot{M}_\bullet = 10^{-3}
\dot{M}_{\bullet,\rm Bondi}$, the lower to $\dot{M}_\bullet =10^{-2}
\dot{M}_{\bullet,\rm Bondi}$ (the same holds for $\omega$Cen).
\label{Msigma}}
\end{figure}

We estimate upper-limits on the masses of the putative central IMBHs
in these globular clusters from 3$\sigma$ upper limits on the radio
continuum flux. We assume that the putative central black hole
accretes the surrounding matter at a rate between 0.1~\% and 1~\% of
the Bondi accretion rate and take into account the uncertainty on the
distance modulus. For 47Tuc, we find a 3$\sigma$ upper limit for the
mass of the central black hole of $2060 - 670 M_\odot$. For NGC6397,
we derive a 3$\sigma$ upper limit of $1290 - 390 M_\odot$ for the
black hole mass.

Massive elliptical and spiral galaxies seem to adhere closely to a
relation between the mass of the central black hole, $M_\bullet$, and
the central velocity dispersion, $\sigma_{\rm c}$ (see
Fig. \ref{Msigma}). Depending on the sample selection and the way the
linear regression is done, the slope and zero-point of this relation
vary \cite{ff05,tr02}. The existence of this relation is however
generally accepted. Although there is theoretically no obvious reason
why this relation should hold with the same slope also in the mass
regime of globular clusters, with the growth of SMBHs most likely
being driven by the merger histories of their host galaxies, it is
nevertheless tempting to do so. The central velocity dispersion of
47Tuc, $\sigma_0 \approx 12$~km/s \cite{ge95}, corresponds to an IMBH
mass $M_\bullet = 1600\,M_\odot$. The central velocity dispersion of
NGC6397 is very low, $\sigma_0 \approx 5.0$~km/s \cite{m91}, and
corresponds to an IMBH mass $M_\bullet = 50\,M_\odot$.

The upper limit for the mass of the black hole in 47Tuc lies just
below the extrapolated $M_\bullet - \sigma_{\rm c}$ relation of
massive elliptical and spiral galaxies as derived by Tremaine et
al. \shortcite{tr02}. As this concerns a 3$\sigma$ upper limit, this
means that the mass of the central black hole in 47Tuc must be much
lower than predicted by this relation, on the condition, of course,
that the black hole accretes at a rate higher than 0.1~\% of the Bondi
rate. It also excluded the presence of a IMBH with a mass of about
0.1~\% of the cluster mass. The extrapolated $M_\bullet - \sigma_{\rm
c}$ relation as derived by Ferrarese \& Ford \shortcite{ff05} predicts
lower $M_\bullet$-values in the globular-cluster regime than the
Tremaine et al. \shortcite{tr02} version and is potentially consistent
with our results. The upper limit for the mass of the black hole in
NGC6397 lies above both renditions of the $M_\bullet - \sigma_{\rm c}$
relation and thus puts no extra constraints on this relation in the
globular-cluster regime. Given the uncertainty concerning the
detection of a black hole in M15 \cite{ge02,m03,ma05}, this means that
no secure detection of an IMBH exists in the low-mass regime
($\sigma_{\rm c} < 15$~km/s) of the $M_\bullet - \sigma_{\rm c}$
relation. Radio-continuum observations of $\omega$Cen yielded a
3$\sigma$ upper limit of $M_\bullet = 1470 - 460$, depending on the
accretion rate \cite{ma05}. We adopt a distance of $4.8 \pm 0.3$~kpc
and a central velocity dispersion $18 \pm 2$~km/s
\cite{vdv06}. Dynamical modeling of the central regions of the Local
Group galaxies NGC205 \cite{v05} and M33 \cite{m01,ge01p} has yielded
no conclusive evidence for IMBHs in the $\sigma_{\rm c} \sim 30$~km/s
regime (with M33, in fact, well below the 1$\sigma$ uncertainties of
the extrapolated $M_\bullet - \sigma_{\rm c}$ relations). This leaves
G1 as the only object with a relatively secure detection of a
sub-$10^6\,M_\odot$ IMBH \cite{geb02,b03,ge05} and that agrees with
the extrapolated $M_\bullet - \sigma_{\rm c}$ relation.

\section{Conclusions} \label{conc}

Using radio-continuum observations with the ATCA radio telescope
array, we estimate the 3$\sigma$ upper-limits on the masses of the
putative central IMBHs in two nearby Galactic globular clusters, 47Tuc
and NGC6397, at respectively $2060 - 670 M_\odot$ and $1290 - 390
M_\odot$. These mass estimates have been derived from 3$\sigma$ upper
limits on the radio continuum flux. We assume that the putative
central black hole accretes the surrounding matter at a rate between
0.1~\% and 1~\% of the Bondi accretion rate and take into account the
uncertainty on the distance modulus.

Black hole masses estimated using radio-continuum observations by
necessity depend on assumptions regarding the density and temperature
of the interstellar medium, and on the accretion rate. Observations of
a few globular clusters provide ``typical'' values for the
characteristics of the interstellar medium. The black hole accretion
rate can be constrained by observations of the central black hole of
the Milky Way and by theoretical models of accretion disks. Still,
radio-continuum estimates of black hole masses probably cannot boast
the same degree of accuracy as mass estimates from e.g. dynamical
models. However, while dynamical models fitted to radial velocities
and proper motions of stars near the center of a globular cluster
yield the most accurate mass estimates, their spatial resolution is
not high enough to identify the central dark mass as a black hole. If,
on the other hand, one detects radio emission coming from accreted
matter falling into a steep gravitational well then one can be sure
that the central object must be very compact, i.e. a black hole. A
cluster of white dwarfs or neutron stars would not be able to generate
such a feature.  In a way, both methods are complementary. Still, it
seems telling that we obtain a 3$\sigma$ upper-limit for the mass of
the putative central black hole in 47Tuc, a nearby massive globular
cluster, that places this object marginally below the extrapolation of
the $M_\bullet - \sigma_{\rm c}$ relation of the bright ellipticals
and spirals. The data for NGC6397 unfortunately do not constrain the
low-mass end of the $M_\bullet - \sigma_{\rm c}$ relation.

Our results hopefully form the basis for further attempts to detect
IMBHs in these globular clusters, which, in the case of a detection,
would lead to a better understanding of the low-mass tail of the
black-hole population and of the accretion rates of central IMBHs in
globular clusters.

\section*{Acknowledgments}
We would like to thank Bob Sault for allowing these observations to be
executed during director's green time. Sven De Rijcke thanks the Fund
for Scientific Research$-$Flanders (Belgium) for approving a travel
grant. Pieter Buyle wishes to thank the Bijzonder OnderzoeksFonds
(BOF) of Ghent University for financial support. We thank the referee,
Tom Maccarone, for his constructive remarks.

\bsp \label{lastpage} 
\begin{thebibliography}{99}
\bibitem[\protect\citename{Andreuzzi et al. }2004]{a04} Andreuzzi, G.,
Testa, V., Marconi, G., Alcaino, G., Alvarado, F., Buonanno, R., 2004,
A\&A, 425, 509
\bibitem[\protect\citename{Baumgardt et al }2003]{b03} Baumgardt, H.,
Makino, J., Hut, P., McMillan, S., Portegies Zwart, S., 2003, ApJ,
589, L25
\bibitem[\protect\citename{Conway, Cornwell, Wilkinson }1990]{ccw90} Conway,
J.E., Cornwell, T.J., Wilkinson, P.N., 1990, MNRAS, 246, 490
\bibitem[\protect\citename{Di Matteo, Carilli, Fabian }2001]{di01} Di Matteo,
T., Carilli, C. L., Fabian, A. C., 2001, ApJ, 547, 731
\bibitem[\protect\citename{Drukier \& Bailyn }2003]{dr03} Drukier,
G. A. \& Bailyn, C. D., 2003, ApJ, 597, L125
\bibitem[\protect\citename{Falcke, K\"ording, Markoff }2004]{fa04}
Falcke, H., K\"ording, E., Markoff, S., 2004, A\&A, 414, 895
\bibitem[\protect\citename{Fender et al. }2003]{fe03} Fender, R. P.,
Gallo, E., Jonker, P., 2003, MNRAS, 343, L99
\bibitem[\protect\citename{Ferrarese \& Ford }2005]{ff05} Ferrarese,
L. \& Ford, H., 2005, SSRv, 116, 523
\bibitem[\protect\citename{Freire et al. }2001]{fr01} Freire, P. C.,
Kramer, M., Lyne, A. G., Camilo, F., Manchester, R. N., D'Amico, N.,
2001, ApJ, 557, L105
\bibitem[\protect\citename{Gebhardt \& Fischer }1995]{ge95} Gebhardt,
K. \& Fischer, P., 1995, AJ, 109, 209
\bibitem[\protect\citename{Gebhardt et al. }1995]{ge95p} Gebhardt, K.,
Pryor, C., Williams, T. B., Hesser, J. E., 1995, AJ, 110, 1699
\bibitem[\protect\citename{Gebhardt et al. }2001]{ge01p} Gebhardt, K.,
Lauer, T. R., Kormendy, J., Pinkney, J., Bower, G. A., Green, R.,
Gull, T., Hutchings, J. B., 2001, AJ, 122, 2469
\bibitem[\protect\citename{Gebhardt, Rich, Ho }2005]{ge05} Gebhardt,
K. G., Rich, R. M., Ho, L. C., 2005, ApJ, 634, 1093
\bibitem[\protect\citename{Gebhardt, Rich, Ho. }2002]{geb02} Gebhardt,
K. G., Rich, R. M., Ho, L. C., 2002, ApJ, 578, L41
\bibitem[\protect\citename{Gerrsen et al. }2002]{ge02} Gerssen, J.,
van der Marel, R. P., Gebhardt, K., Guhathakurta, P., Peterson, R C.,
Pryor, C., 2002, AJ, 124, 3270
\bibitem[\protect\citename{Gratton et al. }2003]{gr03} Gratton, R. G.,
Bragaglia, A., Carretta, E., Clementini, G. Desidera, S., Grundahl,
F., Lucatello, S., A\&A, 408, 529
\bibitem[\protect\citename{Howell, Guhathakurta, Gilliland }2000]{ho00} Howell, J. H.,
Guhathakurta, P., Gilliland, R. L., 2000, PASP, 112, 1200
\bibitem[\protect\citename{Kawakatu et al. }2003]{ka03} Kawakatu, N.,
Umemura, M., Mori, M., 2003, ApJ, 583, 85
\bibitem[\protect\citename{Kawakatu \& Umemura }2005]{no05} Kawakatu,
N. \& Umemura, M., 2005, ApJ, 628, 721
\bibitem[\protect\citename{Krumholz, McKee, Klein }2005]{kr05}
Krumholz, M. R., McKee, C. F., Klein, R. I., 2005, ApJ, 618, 757
\bibitem[\protect\citename{Maccarone }2004]{ma04} Maccarone, T. J.,
2004, MNRAS, 351, 1049
\bibitem[\protect\citename{Maccarone, Fender, Tzioumis }2005]{ma05} Maccarone,
T. J., Fender, R. P., Tzioumis, A. K., 2005, MNRAS, 456L, 17
\bibitem[\protect\citename{Madau \& Rees }2001]{ma01} Madau, P. \&
Rees, M. J., 2001, ApJ, 551, L27
\bibitem[\protect\citename{McConnel et al. }2004]{mc04} McConnell, D.,
Deshpande, A. A., Connors, T., Ables, J. G., 2004, MNRAS, 348, 1409
\bibitem[\protect\citename{McNamara, Harrison, Anderson }2003]{m03}
McNamara, B. J., Harrison, T. E., Anderson, J., 2003, ApJ, 595, 187
\bibitem[\protect\citename{Merloni, Heinz, di Matteo }2003]{mhd03}
Merloni, A., Heinz, S., di Matteo, T., 2003, MNRAS, 345, 1057
\bibitem[\protect\citename{Merritt, Ferrarese, Joseph }2001]{m01}
Merritt, D., Ferrarese, L., Joseph, C. L., 2001, Science, Vol. 293,
1116
\bibitem[\protect\citename{Meylan \& Mayor }1991]{m91} Meylan, G. \&
Mayor, M., 1991, A\&A, 250, 113
\bibitem[\protect\citename{Miller \& Hamilton }2002]{mi02} Miller,
M. C. \& Hamilton, D. P., 2002, MNRAS, 330, 232
\bibitem[\protect\citename{Perna et al. }2003]{pe03} Perna, R.,
Narayan, R., Rybicki, G., Stella, L., Treves, A., 2003, ApJ, 594, 936
\bibitem[\protect\citename{Portegies Zwart \& McMillan }2002]{pz02}
Portegies Zwart, S. F. \& McMillan, S. L. W., 2002, ApJ, 576, 899
\bibitem[\protect\citename{Proga \& Begelman }2003]{pb03} Proga, D. \&
Begelman, M. C., 2003, ApJ, 582, 69
\bibitem[\protect\citename{Quataert \& Gruzinov }2000]{qg00} Quataert,
E. \& Gruzinov, A., 2000, ApJ, 545, 842
\bibitem[\protect\citename{Reid \& Gizis }1998]{rg98} Reid, I. N.,
Gizis, J. E., 1998, ApJ, 116, 2929
\bibitem[\protect\citename{Sault \& Conway }1999]{sc99} Sault, R.J.
\& Conway, J.E., 1999, Synthesis Imaging in Radio Astronomy II, ASP Conference Series, Vol. 180, eds. Taylor G.B., Carilli C.L. and Perley R.A.
\bibitem[\protect\citename{Sault, Teuben, Wright }1995]{stw95} Sault, R.J.,
Teuben, P.J., Wright, M.C.H., 1995, in Shaw R., Payne H.E., Hayes J.J.E., eds, ASP Conf. Ser. Vol. 77, Astronomical Data Analysis Software and Systems IV. Astron. Soc. Pac., San Fransisco, p. 433
\bibitem[\protect\citename{Tremaine, Gebhardt, Bender }2002]{tr02}
Tremaine, S., Gebhardt, K., Bender, R. et al., 2002, ApJ, 574, 740
\bibitem[\protect\citename{Valluri et al. }2005]{v05} Valluri, M.,
Ferrarese, L., Merritt, D., Joseph, C. L., 2005, ApJ, 628, 137
\bibitem[\protect\citename{van de Ven et al. }2006]{vdv06} van de Ven,
G., van den Bosch, R. C. E., Verolme, E. K., de Zeeuw, P. T., 2006,
A\&A, 445, 513
\end{thebibliography}
\end{document}